# Measurement of ultrashort laser ablation of four metals (Al, Cu, Ni, W) in single pulse regime

T. Genieys[1], M. Sentis[1], O. Utéza[1]

[1]Aix-Marseille University, CNRS, LP3 UMR 7341, F-13288 Marseille, France

*Corresponding author*: uteza@lp3.univ-mrs.fr

**Abstract**

We provide measurements of ablation of four post-transition and transition metals (aluminum, copper, nickel and tungsten) irradiated by single 800 nm laser pulses, in ultrashort regime from 100 fs pulse duration down to 15 fs covering a temporal range little explored yet. For each metal and pulse duration tested, we measure its ablation characteristics (depth and diameter) as a function of incident energy allowing us to determine its laser-induced ablation threshold and ablation rate in single-shot regime. For all metals studied, we observe constant ablation threshold fluence as a function of pulse duration extending this scaling law to pulse duration of few-optical-cycles. We also provide evidence of the interest of adjusting the incident fluence to maximize the energy specific ablation depth but also of the absence of any peculiar advantage related to the use of extremely short pulse duration for ablation purposes. Those informative and detailed ablation data have been obtained in single pulse regime and in air ambiance. They can serve as rewarding feedback for further establishing smart strategy for femtosecond laser micromachining and laser damage handling of metallic and metal-based components as well as for enhancing accuracy of modeling of femtosecond laser interaction with metals in ultrashort regime.

## 1 Introduction

Femtosecond lasers are unique tools to machine materials with minimized thermal budget and high-quality process capability practically free of collateral effects [1,2]. In order to calibrate material transformation upon laser excitation and to develop subsequent micromachining processes, determination of laser-induced ablation threshold fluence is mandatory. Such measurements have been performed on a large variety of materials, including metals, semi-conductors and dielectrics [3-6], and in a very broad range of pulse duration until few femtoseconds. In dielectric materials, initially optically transparent to 800 nm pulses, a strong decrease in the ablation threshold fluence has been measured when the pulse duration is reduced from hundreds of femtoseconds to few optical cycles [3,7,8]. In the case of semiconductors, a much less marked decrease of the ablation threshold fluence has been measured for pulse durations ranging from 5 to 400 fs [9].

In metals, the evolution of the ablation threshold fluence as a function of the pulse duration can be separated into two different regimes. For the pulse durations longer than the electron-phonon coupling time, the ablation threshold fluence increases as a function of the square root of the pulse duration and for pulse durations shorter than the electron-phonon coupling time, the ablation threshold remains constant when the pulse duration decreases [10,11]. This principle has been confirmed for pulse durations down to a few hundred femtoseconds. In femtosecond regime, numerous works have been carried out, but mainly for pulses of duration ≥ 100 fs or in multi-pulse mode [4,6,10,12] conveniently corresponding to today-mature commercial lasers suitable for the development of applications.

In the context of ultrashort pulses (<< 50 fs), one can cite few relevant works. For instance in [4], it was derived from measurements in multi-pulse regime single-shot ablation threshold fluence of copper in vacuum for various pulse durations, including experimental data at 10, 30, 250 and 550 fs. In particular, they demonstrate weak scaling with pulse duration ($F_{th,10fs}$ = 0.77 J/cm² and $F_{th,550fs}$ = 0.95 J/cm²). Other single-shot measurements obtained with Ti:Sapphire fs lasers in air, and including copper, aluminum or tungsten, yield diverse values, e.g. $F_{th,Al}$ = 0.51 J/cm² (33 fs) [13], $F_{th,Cu}$ = 1.7 J/cm² (100 fs) [14], $F_{th,Cu}$ = 0.58 J/cm² (150 fs, however issued from fit of a multi-pulse incubation study) [6], $F_{th,W}$ = 0.44 J/cm² (100 fs) [14]. Nonetheless experimental data are still rare in the ultrashort regime especially down to few-optical-cycle pulse durations (≤ 15 fs) and sometimes showing high controversy, which justifies and motivates further research. Moreover, measurements of ablation and damage thresholds of materials (including metals) are in high demand due to the rapid and impressive developments of high-peak power (PW class) ultrashort-pulse laser systems and associated fascinating applications and fundamental breakthroughs [15], and due to the development of high-quality micromachining processes potentially taking profit of the use of ultrashort pulses [16]. For instance, this immediately calls for laser damage certification of optical components and the determination of irradiation levels for calibrating matter transformation. Another motivation roots in providing detailed ablation data for progressing in fundamental understanding of ultrashort laser interaction with materials and the quest for establishing predictive quantitative models [17,18]. In this context, performing experiments with ultrashort pulses of different pulse duration is of prominent interest allowing us to test ablation phenomenon with pulses shorter than the characteristic energy coupling times of the material, like its electron-phonon coupling time or even its electron-electron equilibration time.

In order to provide measurements of ablation of metals in ultrashort regime, we thus perform specific measurements of laser-matter interaction (see description of experiments in Section 2). We concentrate our study on four post-transition and transition metals (aluminum, copper, nickel and tungsten) of academic and industrial interest. Note that the single-shot mode of interaction is here exclusively studied in order to get rid of any incubation effects and to provide calibrated ablation data on which strategy of micromachining can be further optimally developed. Firstly we evaluate the laser-induced ablation threshold fluence on a wide range of pulse duration (15 – 100 fs, see Section 3.1) and secondly, we measure the evolution of the ablation characteristics (ablated diameter and depth) as a function of the incident fluence (see Section 3.2). This finally helps to provide insights into energy specific ablation depth and to evaluate the interest of few-optical-cycle laser pulses for micromachining of metals.

## 2 Experimental procedure and details

All irradiation experiments are performed in single-shot regime using the beam line 5a of ASUR platform at LP3 laboratory. Nominally, the beam line delivers linearly polarized ~ 30 fs (FWHM) pulses at 100 Hz with 800 nm central wavelength ($\Delta\lambda \cong 760 - 840$ nm FWHM) and a maximum energy of 1 mJ with 1% rms fluctuations. Afterwards, two experimental arrangements have been used (Fig. 1). The test bench A allows changing the pulse duration from 30 to 100 fs by pre-chirping the beam through compressor grating adjustments. The test bench B provides access to shorter pulse durations. Cross-polarized wave generation (XPW) is used to broaden the spectrum (720 – 880 nm) and final compression based on chirped mirrors and a pair of fused silica wedges yields pulses with duration down to 15 fs, and energy up to 30 µJ with fluctuations of 2.5% rms. The pump signal is suppressed by two consecutive pairs of Brewster polarizers. On both test-benches the incident energy is controlled by the combination of a half-wave plate and a set of four thin Brewster polarizers. Another half-wave plate is inserted before the focusing optics to manage the beam polarization.

The experiments are done at normal incidence and using the same incident (parallel) linear beam state of polarization. The beam is focused by a 90° off-axis parabola of 152.4 mm focal length on setup A and of 50.8 mm focal length on setup B. We used a smaller effective focal length (EFL) off-axis parabola on setup B to decrease the sensitivity to Kerr effect when dealing with intense few-optical-cycle pulses [19]. The spatial characterization of the beam is done by imaging the laser focal spot onto a CCD beam analyzer (*Gentec* Beamage) through a microscope objective. The measurement is calibrated with respect to a Ronchi grating of 200 lines per mm. The two beams have a Gaussian spatial distribution and propagation, and their radius at $1/e^2$ in the focal plane are determined before each experimental test. Measurement of pulse duration is provided by a second-order interferometric autocorrelator (Femtometer, *Femtolaser*) for pulse durations of 15 and 30 fs, and by a single-shot autocorrelator (Bonsai, *Amplitude Technologies*) for pulse durations of 50 and 100 fs. This measurement was done systematically before the focusing parabola, considering all the dispersive optics present on the beam path before it. The target surface is positioned normal to the laser beam at the focal plane using repeated z-scan procedures at decreasing incident energies and with in-situ imaging diagnostic. The sample is placed in the focal plane with a precision of 50 µm, much less than the Rayleigh length of the beams on the two setups.

During experiments the intensity of the beam is varied up to ~ $4.10^{14}$ W/cm² and the peak power up to ~ GW. To evaluate the importance of self-focusing induced by optical Kerr nonlinearity and of defocusing related to self-produced plasma in air before the target, we measure in details the beam propagation as a function of the incident energy. On setup A at 30 fs, we measure a spatial shift of the focal plane position towards the focusing optics from the incident energy of ~ 10 µJ. On setup B, nonlinear effects in air start to be significant from the incident energy of 4.3 µJ [19]. When fitting the experimental data, for instance for the determination of the ablation threshold, we thus exclude the measurement points obtained at higher incident energy because of the progressive loss of control of the incident intensity on target and subsequent risks of biased interpretation.

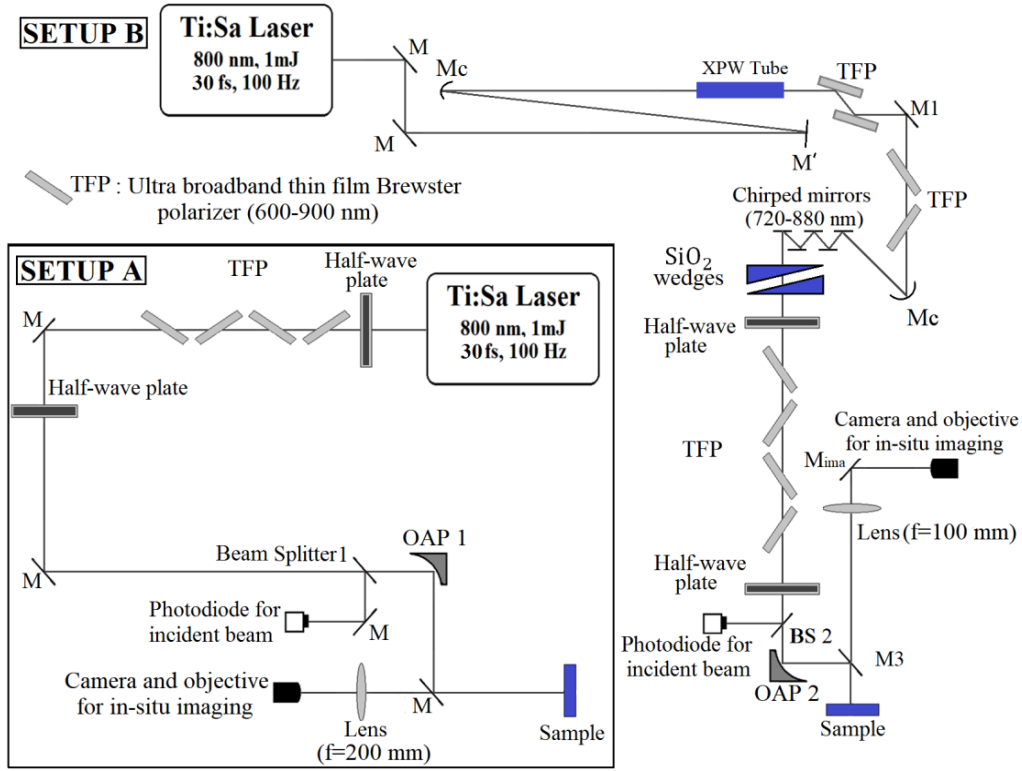

**Figure 1** Experimental test-benches. **Setup A** (experiments at 30, 50 and 100 fs): M: low dispersion 45° incidence dielectric mirrors (750 – 850 nm), OAP1: Silver-coated off-axis parabola (effective focal length EFL = 152.4 mm), Beam splitter 1: broadband 70% transmission / 30% reflection (p-polarization) at 45° for the wavelength range 700 nm - 950 nm. **Setup B** (15 fs): M: same as for setup A; M': low dispersion 0° incidence dielectric mirror (750 – 850 nm); M1: low dispersion Ag mirror; Mc: metallic-coated concave mirror (5 m radius of curvature) for pump beam focusing and XPW beam collimation; XPW tube with a hollow core fiber for pump spatial filtering and two $BaF_2$ nonlinear crystals for spectrum enlargement; BS2: broadband (700 – 950 nm) 70% transmission / 30% reflection (p-polarization) at 45°; OAP2: Gold-coated off-axis parabola (EFL = 50.8 mm); M3: dielectric mirror with reflectivity of 67% at 45° for 720 – 880 nm wavelength range.

The four metals studied were supplied by *Goodfellow Inc.* and have a purity of more than 99.9%. Their thickness varies between 0.5 and 3.2 mm, and the roughness (*Ra* parameter) of the materials was measured using an atomic force microscope (*PSIA* XE-100). Those characteristics as well as their main thermo-physical properties are summarized in the table 1. It is important to note that the surface finish of the sample influences the sample reflectivity. The reflectivity $R_{measured}$ scattered by a real rough surface (with *Ra* characteristics) can be linked to the reflectivity of a perfectly smooth surface $R_{material}$ by the formula: $R_{measured} \approx R_{material} \exp\left|-\left(\frac{4\pi Ra}{\lambda}\right)^2\right|$, (eq. 1) [20]. The reflectivity $R_{measured}$ of the different metals was measured using the collimated beam line 5a operated at very low incident energy (fluence) to provide a measurement at 800 nm (and with the corresponding spectral width) for the samples without inducing any change of their properties (unperturbed material). In those conditions in which the detector is placed at a large distance of the sample, the monitored reflectivity ($R_{measured}$) does not incorporate the diffuse part reflected by the sample (especially for Al and Cu having high *Ra* characteristics). The values of the reflectivity coefficient obtained from those measurements are in very good agreement with the literature values obtained at low energy with continuous beams [21,22] and with the optical response of the material given by

Fresnel's formulas (and corresponding to the reflectivity $R_{calculated}$ using coefficients of tabulated dielectric function or with the reflectivity $R_{material}$ of the perfectly smooth surface when using Eq. 1) (see Table 1).

| Material | Aluminum | Copper | Nickel | Tungsten |
|---|---|---|---|---|
| Electronic configuration | [Ne] $3s^23p^1$ | [Ar] $3d^{10}4s^1$ | [Ar] $3d^84s^2$ | [Xe] $4f^{14}5d^46s^2$ |
| Molar mass $M$ (g.mol$^{-1}$) | 26.98 | 63.55 | 58.69 | 183.84 |
| Atomic density $n_{at}$ (m$^{-3}$) | 6.02×10$^{28}$ | 8.49×10$^{28}$ | 9.13×10$^{28}$ | 6.32×10$^{28}$ |
| Number of free electrons per atom | 3 | 1 | 2 | 2 |
| Volumetric mass density, $\rho$ (g.cm$^{-3}$) | 2.70 | 8.96 | 8.90 | 19.3 |
| Fermi energy (eV) and temperature ($T_F = E_F/k_B$) (K) | 10.8 - 125217 | 7.05 - 81739 | 11.7 - 135652 | 9.2 - 106667 |
| Electron work function*, $E_{esc}$ (eV) | 4.17 | 4.76 | 5.2 | 4.55 |
| Atomic bond dissociation energy, $E_{bd}$ (eV) | 3.435 | 3.5 | 4.465 | 8.835 |
| Melting temperature, $T_m$ (K) | 933 | 1357 | 1728 | 3695 |
| Vaporization temperature (K) | 2792 | 2835 | 3186 | 5826 |
| Enthalpy of melting (kJ/mol) | 10.79 | 13.05 | 17.48 | 35.4 |
| Enthalpy of vaporization (kJ/mol) | 294.0 | 300.3 | 370.4 | 824.0 |
| Sum of enthalpy of melting and vaporization $\Omega$ (J/mm$^3$) | 30.48 | 44.07 | 58.86 | 90.75 |
| Specific heat capacity, $C_p$ (J/g.K) | 0.897 | 0.38 | 0.44 | 0.13 |
| Electron thermal conductivity $k_e$ (W.m$^{-1}$.K$^{-1}$) | 237 | 401 | 90,7 | 174 |
| Electron heat capacity (J/m$^3$.K) (300 K) ($C_e \cong \pi^2 k_B n_e(\frac{T_e}{T_F})/2$) | 29240 | 21230 | 27541 | 24220 |
| Plasma frequency $\omega_p$ (s$^{-1}$) | 2.395×10$^{16}$ | 2.84×10$^{16}$ | 2.95×10$^{16}$ | 2.45×10$^{16}$ |
| Index of refraction, $n + ik$ | 2.7673 + i8.3543 [21] | 0.25352 + i5.0131 [25] | 2.48 + i4.455 [26] | 3.6528 + i2.6976 [26] |
| Thickness (mm) | 0.5 | 1 | 3.2 | 2 |
| Sample dimensions (mm) | 25 x 25 | 10 x 10 | 25 x 25 | 10 x 10 |
| $Ra$ (nm) | 20 | 17 | 5 | 8 |
| Reflectivity $R_{measured}$ | 0.773 | 0.908 | 0.682 | 0.497 |
| Reflectivity of a perfectly smooth surface, $R_{material}$ (as deduced from Eq. 1) | 0.853 | 0.975 | 0.686 | 0.505 |
| Reflectivity $R_{calculated}$ | 0.868 | 0.962 | 0.69 | 0.495 |

**Table 1** Thermo-physical and optical properties of studied metals [22-24] and physical characteristics and optical properties measured on the samples (in blue rectangles). For 800 nm – 1.55 eV when relevant. * For the electron work function, the value is averaged between the different faces of a mono-crystalline sample.

## 3 Results and discussion

### 3.1 Laser-induced ablation threshold fluence in ultrashort regime

To determine the ablation threshold fluence, the diameter-regression technique is used considering a Gaussian spatial beam distribution and evaluation of the interaction of the laser with the metallic samples at different levels of incident energy using confocal optical microscopy [27,28] (see illustration in case of copper in Fig. 2). The ablation threshold fluence $F_{th}$ is expressed as the laser peak fluence $F_{th}=\frac{2E_{th}}{\pi \omega^2_0}$ with $E_{th}$ the measured incident energy for which the ablated diameter is equal to zero and $\omega_0$ the radius of the focal spot at $1/e^2$ determined using beam analyzer (see section 2 and quantitative values given in table 2). It is important to note that nearly the same ablation threshold values (difference much inferior to measurement and fit uncertainties) are determined when considering the beam waist inferred from the slope of diameter regression curves. The threshold values obtained for each pulse duration and metals are shown in table 2.

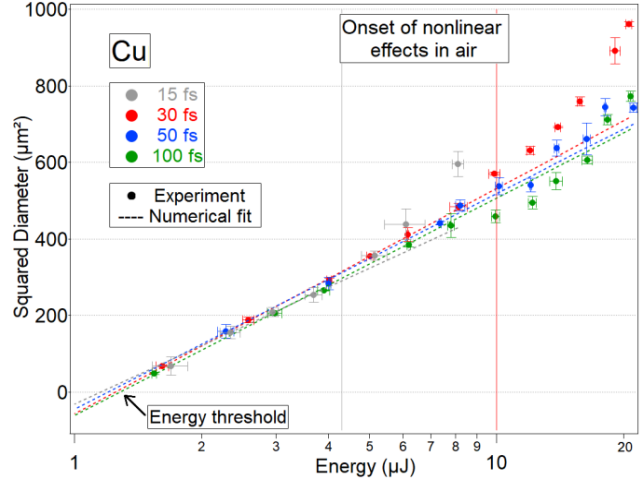

**Figure 2** Evolution of squared diameter $D^2$ ($D^2 = 2\omega_0^2 \ln(E/E_{th})$, [27]) versus incident energy for each pulse duration (illustrated for copper). The threshold is determined by the fit energy value for which $D^2 = 0$. Each dot is an average value of 8 measurements, with error bars being the standard deviation. The horizontal error bars correspond to shot-to-shot fluctuations measured on a photodiode. The grey and red vertical solid lines locate the onset of nonlinear effects in air at 15 and 30 fs as measured on the two setups.

| Pulse Duration | 15 fs | 30 fs | 50 fs | 100 fs | Scaling law (Eq. 2) | Scaling law (Eq. 3) |
|---|---|---|---|---|---|---|
| Aluminum, $F_{th}$ (J/cm²) | 0.232 | 0.239 | 0.240 | 0.229 | 0.285 | 0.18 |
| Copper, $F_{th}$ (J/cm²) | 0.636 | 0.651 | 0.637 | 0.659 | 4.28 | 2.425 |
| Nickel, $F_{th}$ (J/cm²) | 0.328 | 0.331 | 0.329 | 0.316 | 0.48 | 0.27 |
| Tungsten, $F_{th}$ (J/cm²) | 0.521 | 0.541 | 0.530 | 0.531 | 0.485 | 0.47 |

**Table 2** Ablation threshold fluence of the four metals, for 15, 30, 50 and 100 fs pulse duration deduced from the diameter regression technique. The waists were measured before the experiments as specified in section 2. For 15 fs: $\omega_0 = 7.75$ µm in case of nickel and $\omega_0 = 10.20$ µm for the other metals, for 30-100 fs and all metals: $\omega_0 = 11$ µm.

For all metals tested, the ablation threshold fluence is shown constant over the pulse duration investigated (15 – 100 fs). Indeed, the small variations observed at different pulse durations are below the error bar ($\leq \cong 0.02$ J/cm² in our experimental work) attached to the uncertainty in $F_{th}$ threshold determination. Such constant behavior of ablation threshold in short pulse duration domain was already mentioned in picosecond and sub-picosecond regime for decades (see for instance [29]) and more recently for femtosecond pulses (see for instance [14]), but it was until now not completely verified for single femtosecond pulses down to few-optical-cycle laser duration ($\cong 15$ fs) providing novelty and relevance to our works. Moreover and interestingly some deviations were reported for copper as in [4] with significant difference observed for very different pulse duration spanning from 10 to 550 fs. However those experiments were not conducted in single-shot regime and considering the ultrashort pulse range (10 and 30 fs cases), the inferred ablation threshold fluence was within few percent difference ($\leq 3\%$) so not really depending on pulse duration as we observed in our experiments.

In order to benchmark our ablation threshold data, and to test the applicability of simplified theoretical approaches, we further compare them to two scaling laws available from literature. A first scaling formula for laser-induced ablation threshold fluence has been proposed by E. Gamaly et al. for femtosecond pulses [30]. It is built on a description based on charge separation of the ablation phenomenon in which the electron must gain during the laser pulse enough energy first to escape the surface (electron work function $E_{esc}$) and secondly to drag the parent ion out of the solid (bond dissociation energy $E_{bd}$). Mathematically, this energetic condition to yield atom removal (ablation) in the surface layer in which absorption takes place is expressed by [30]:

$$F_{th,elec} = \frac{3}{4}(E_{bd} + E_{esc})\frac{l_{opd} n_e}{A} \quad \text{(Eq. 2)},$$

where $A$ is the absorption. In this calculation (see results in table 2), the absorption is inferred from the relation $A = 1 - R$, with $R = R_{material}$, the reflectivity corresponding to the perfect (smooth) material (see table 1); and the free electron density is taken equal to the atom density based on the consideration that the above condition is formulated for an individual free electron gaining enough energy to remove its parent ion. The last parameter corresponds to the optical penetration depth $l_{opd} = \frac{c}{2\omega k}$, $c$, $\omega$ and $k$ being respectively the speed of light, the laser frequency (2.355×10$^{-15}$ s$^{-1}$ at 800 nm) and the extinction coefficient available in table 1. It is interesting to note that Eq.2 does not explicitly depend on pulse duration, which is in very good agreement with our experimental results showing that a constant ablation threshold is obtained on the whole pulse duration range tested (15 – 100 fs). From a qualitative point of view, the relative comparison of the measured and calculated ablation threshold fluences is correctly restored for all metals tested. Quantitatively, we observe a fair correspondence for aluminum, nickel and tungsten and not at all for copper for which strong divergence is observed. For copper, which has a reflectivity coefficient close to 1, the calculation of the ablation threshold is highly sensitive to small variations of absorption. Uncertainty about the exact value of this parameter leads to possibly large discrepancy between the scaling equation and the measurement. Finally, the physical concept of creation of an electric field resulting from electron charge separation to yield ablation still forms an open debate because of no really convincing observation of such phenomenon until now. In addition, it is especially questionable for metals for which high electron mobility and rapid charge neutralization is expected.

Another approach based on equilibrium thermodynamic considerations was also developed (eq. 3) [31]. It expresses that the energy equal to the energy necessary to melt and further evaporate the atoms in the surface layer should be transferred to the lattice (without any considerations of transport and diffusive losses out of the focal volume):

$$F_{th,evap} = \frac{\Omega(in\frac{J}{cm^3}) l_{opd}}{A} \quad (3)$$

Where $\Omega$ is the sum of the enthalpy of melting and vaporization and of the energy necessary to raise the initial temperature of the sample ($T_0 = 293$ K) to its melting temperature ($\Delta E = \rho C_p (T_m - T_0)$. The results are listed in table 2 for comparison. We observe a fair agreement, with a relative comparison of the four metals correctly restored. As for the other scaling law

(2), a large discrepancy is calculated for cooper, which we again attribute to the high sensitivity of the result to the exact value of the absorption parameter.

As a first short summary, both equations allow a correct evaluation of the metal ablation threshold fluence, except in the case of copper nonetheless which has a reflectivity close to unity. However, for more accurate comparison with experimental results and further progressing towards full predictability, more detailed experiments (collecting in particular time-resolved data) to provide a better estimation of the absorption parameter and detailed theoretical developments are necessary. This is in agreement with other works showing that other mechanisms are at play when ablating matter in femtosecond regime, including not classical boiling (as before considered in Eq. 3) but phase explosion or spallation effects [32,33] and involving reaching transient critical lattice temperature or stress.

*3.2 Ablation characteristics as a function of incident energy*

The ablation diameters and depths have been measured using confocal optical microscopy. Their evolution as a function of incident fluence is shown in Figure 3 for the four metals at pulse durations of 15, 30, 50 and 100 fs. Considering the evolution of ablation diameters, the trend line based on the Gaussian spatial distribution of the laser beam and the deterministic nature of the interaction as followed in [27,28] is added to the data. The numerical data are obtained considering the beam waist measured experimentally (section 2 and legend of table 2 for the quantitative values). We note the excellent agreement of the numerical and experimental data. This holds true for most of the fluence range tested, except for the points at relatively high fluence which whatever the target departs in the same way (in other words for setup A from $F > \sim 8$ J/cm²) from the trend line. This is not surprising because those high fluences significantly exceed the energetic levels for which we measured the onset of nonlinear effects on both setups. Due to self-focusing and air ionization in front of the target, the beam propagation is distorted with reshaping of its space and time modal distribution yielding to beam enlargement and limitation of the beam fluence on target [34], consistently with the observation of larger ablated diameters at very high incident fluence. Only small differences can be seen depending on pulse duration (with higher sensitivity in case of shorter pulses, see figure 3-right) because of the slight variations of nonlinear index and air ionization properties with that parameter in the considered range. This is consistent with measurements of nonlinear propagation in air showing a lower energy value (but only by a factor 0.75) for inducing filamentation when comparing 40 and 125 fs pulses [35].

The interpretation is less straightforward when considering the evolution of the ablated depth. This observable ($d_{ablated}$) is strongly related to the evolution of the penetration depth of the laser radiation and of the electron heat conduction with applied fluence which are two parameters that are changing dramatically when the excitation is intense enough. When studying the ablation rate per pulse in multi-pulse regime [14,36,37], it was shown that two regimes of ablation can be observed: a first regime at small fluences where the ablation rate is small (so called "gentle ablation") with a logarithmic dependence ($L = l_{opd} \, ln\left(\frac{F}{F_{th}}\right)$) and its characteristic slope parameter related to the optical penetration depth ($l_{opd} = 1/\alpha = l_s/2$, with absorption coefficient $\alpha$ (based on intensity) and skin depth $l_s$); and a second regime for higher fluences (referred as "strong ablation") also characterized by a logarithmic dependence ($L = l_{heat} \, ln\left(\frac{F}{F_{th}}\right)$) but with a much higher rate and the slope corresponding to the effective electronic heat penetration depth $l_{heat}$. In our experiments, the experimental configuration of the study is different because it addresses the single-shot regime only. Importantly, this will allow us to rule out any influence of the shot-to-shot evolution of the material absorption (as it is the case in multi-shot regime) and to provide a precise evaluation and explanation of the most probable physical mechanism at play in the evolution of the ablated depth, especially at high fluences. However, we qualitatively retrieve similar evolution (see figure 3) with large ablated depths at high fluences (except tungsten nonetheless). Indeed, dramatic changes of the absorptivity during the pulse and of the electron heat conductivity and electron-phonon coupling during and after the pulse also occur when the excitation is intense enough, providing reason to the observed evolution of the ablation depth with applied fluence [18,38-43]. Qualitatively, they are similar to changes imprinted in the material pulse after pulse (incubation effects) which are characterized by progressive modification of its absorptivity and of its initial properties in general. This is for instance evidenced by smaller ablation thresholds measured in multi-pulse regime as compared to the single-shot value [4].

Returning to the analysis of our single-shot experiment, we thus plot on figure 3 the two logarithmic dependence curves of the ablated depth as a function of applied fluence ($d = l_{eff} \, ln\left(\frac{F}{F_{th}}\right)$, with successively $l_{eff} = l_s$, the skin depth, and $l_{heat}$ the electronic heat penetration depth). The quantity $l_s$ is calculated from the formula, $l_s = \frac{c}{\omega k}$, using the tabulated extinction coefficient (see table 1). This is a reasonable approximation around the ablation threshold becoming much more severe at higher fluences when the transient optical properties are changed upon laser energy coupling in the material [18,40].

We calculate $l_{heat}$ from the following equation: $l_{heat} = \sqrt{D \tau_{diff,heat}}$, where $D$ is the electron thermal diffusivity ($D = k_e/C_e$, with $k_e$ the electron thermal conductivity and $C_e$ the electron heat capacity). In first approximation, we do not consider the variations of $k_e$ and $C_e$ which are dependent of the space and time evolution of the electron temperature. We take them constant corresponding to equilibrium conditions before excitation (see table 1). This hypothesis is severe in the skin depth volume where the laser energy is deposited and the excitation (electron temperature) intense but however more acceptable when considering the complete heated depth volume where the electron and lattice temperature are much smaller than at the surface of the material. As electrons are the essential vector of energy transport in metals, the time $\tau_{diff,heat}$ basically corresponds to the duration during which the deposited energy has time to diffuse before to be communicated to the lattice. So, conveniently we take: $\tau_{diff,heat} = \tau_{ei}$, where $\tau_{ei}$ is the characteristic time for energy transfer to the lattice of the energy transiently stored in the electron sub-system. It can be approximated by the following formula: $\tau_{ei} \approx \frac{M_{atom}}{m_e \nu_{ei}}$ [30], with $M_{atom}$, the atomic mass of the metal considered, $m_e$ the free electron mass and $\nu_{ei}$ the electron-ion collision frequency. The quantity $\nu_{ei}$ varies with excitation, however we assume that in condition of ablation the electron-ion collision frequency almost coincides with the plasma frequency $\nu_{ei} \cong \omega_p$ [30] ($\omega_p$ being the plasma frequency).

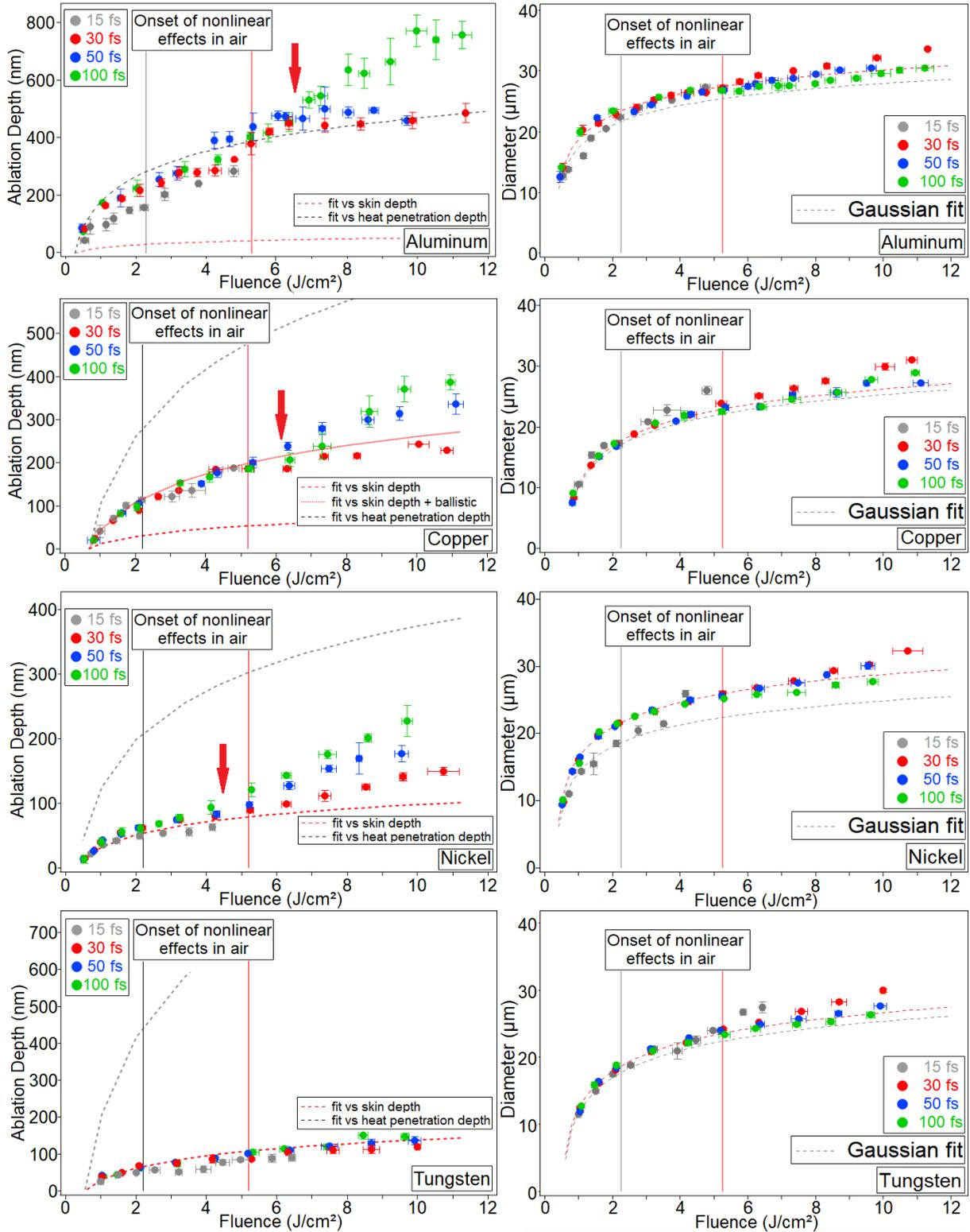

**Figure 3** Evolution of ablated depth (left) and diameter (right) as a function of fluence for the four pulse durations studied. Each point is averaged on eight experiments. (Right) The dotted curves are trend lines based on the Gaussian spatial distribution of the laser beam and the deterministic approach as followed in [27] and taking $w_0 = w_{0,imagery}$ (see equation in legend of figure 2). (Left) The dotted curves correspond to an equation of material removal based on the optical (in red) and electron heat (in grey) penetration depth. When relevant, a third fit (dot red) including consideration of ballistic electrons is added (for Al and Cu) and red arrows (for Al, Cu and Ni) indicate the fluence level from which the ablated depth becomes pulse-duration dependent.

The parameters $l_s$ and $l_{heat}$ are listed for each material (see table 3) and the corresponding fit curves are visible in Figure 3. In the calculation of these parameters, we have not included any dependency with respect to the pulse duration in accordance with the experimental results. Indeed, the evolution of the ablation depths is the same independently of the pulse duration up to an incident fluence level (for instance $F \cong 20\ F_{th} \cong 4.7$ J/cm² in the case of Al, $F \cong 10\ F_{th} \cong 6.5$ J/cm² in the case of Cu and no pulse-duration-dependent deviation until the maximum fluence tested for W). The fluence from which a deviation is measured is different depending on the metal studied.

| Material | Aluminum | Copper | Nickel | Tungsten |
|---|---|---|---|---|
| First log-dependence fit (skin depth), $l_s$ (nm) | 15.25 | 25.4 | 28.6 | 47.2 |
| Second log-dependence fit (electronic heat penetration depth), $l_{heat}$ (nm) | 129 | 277 | 109 | 313 |
| Electron-ion transfer time $\tau_{ei}$ (ps) | 2.21 | 4.08 | 3.63 | 13.6 |

**Table 3** Slope parameters $l_s$ and $l_{heat}$ for all metals and used for plotting the fit curves in figure 3. The electron-ion transfer time $\tau_{ei}$ entering in the calculation of $l_{heat}$ is also listed for information.

Considering figure 3, we observe that the experimental evolution of the ablated depth is in excellent agreement with the fit issued from skin depth scaling for tungsten in the whole energy range tested ($F_{max} \sim 10$ J/cm²) and for nickel until approximately the fluence of 15 $F_{th}$ (~ 5 J/cm²). These two materials have small electron thermal conductivity and our results indicate that electron energy transport does not play a significant role at least in the incident energy range tested (for Ni, ≤ 5 J/cm² nonetheless), the ablated volume corresponding to the one in which the laser energy is coupled in. In contrast a good agreement is obtained considering the scaling based on the heat penetration depth for aluminum until ~ 30 $F_{th}$ (~ 7 J/cm²) while no real satisfactory agreement is observed for copper whatever the scaling considered. For that last reason, we add another fit in case of copper in which the parameter $l_{eff}$ is left free to provide correspondence with the depths experimentally measured (see figure 3). Interestingly, an excellent agreement is obtained for $l_{eff}$ = 95 nm which corresponds to the sum of the skin depth and the range on which ballistic electrons travel in Cu as it was estimated for femtosecond pulses [44]. This suggests the significant role played by ballistic electrons to contribute to excite a volume much superior to the focal volume [44,45]. Nonetheless we interpret those observations made on aluminum and copper as an ablation process progressively being strongly dominated by electron transport (especially Cu) and thermal diffusion (especially Al) mechanisms, even here extended to pulses of few-optical-cycle pulse duration ($\cong$ 15 fs). Moreover, among the four metals tested here, both Al and Cu have the highest initial electron thermal conductivity and the smallest enthalpy and temperature of transformation (phase transition) (see table 1) making more accessible the removal of large depth and ablated volume (as shown in figure 3).

For Al, Cu and Ni, a strong deviation with higher ablation rate can be remarked (see the red arrows in figure 3) from an eventual saturation effect at high incident fluences (as for instance expected from the fits). Note that this deviation is dependent of the pulse duration, as larger ablation depths are measured for longer pulse durations. As examples, those deviations occur from $F \sim 7$ J/cm² for Al and Cu, from $F \sim 5$ J/cm² for Ni and no such effect is observed for W for which ablation is dominated by the laser penetration depth only in the whole energy range tested. We discard here any significant influence of the nonlinear effects developing in air in front of the target because the observed deviations take place at different incident fluences for each metal (from $F \sim 5$ J/cm² for nickel and no deviation until $F \sim 10$ J/cm² for tungsten for the two extreme cases). Moreover, the nonlinear effects developing in air in front of the target would yield a limitation of the fluence at its surface which is not in favor of obtaining even higher ablated depths as it is observed in figure 3. We rather attribute this behavior to the complex changes experienced by electron transport (electron thermal conductivity) when excitation is well above the ablation threshold [36,38,41]. Indeed, it was established the scaling of the electron thermal conductivity with electron and lattice temperature under the following form [18,38,46] (expression valid for electron temperature up to the Fermi temperature $T_F=E_F/k_b$, which is a reasonable assumption here): $k_e \propto \frac{T_e}{T_e^2+T_i}$. The lattice temperature $T_i$ is expected not to vary much depending on pulse duration. This is supported by macroscopic ablation observables in favor of similar energy deposition whatever the whole pulse duration tested (15 – 100 fs) as the measurement of a constant ablation threshold and similar ablation characteristics on a large fluence range above the ablation threshold. Moreover, at the time scale of laser energy deposition and further electron transport (ps), the lattice temperature does not raise much because the energy transfer to the lattice is far from being completed. So the role of the lattice temperature can be neglected in first approximation in the previous scaling. Maximum electron temperature being smaller for longer pulses (providing smaller heating rate), it is thus expected a facilitated electron transport deep into the target compared to shorter pulses ($D = k_e/C_e \propto (1/T_e)/T_e = 1/T_e^2$). This finally provides an energy coupling and dissipation (further ablated volume) in a larger volume (much superior to the focal volume) at long pulse duration. Then this effect becomes definitely apparent at high fluences when conditions to yield high ablated volume (phase explosion) are met (as qualitatively indicated by the arrows in figure 3) [33,47]. As a supporting observation note that the most dramatic changes are again observed for the two materials (Al and Cu) being the most sensitive to energy transport and thermal processes in first approximation (high electron thermal conductivity and small enthalpy and temperature of transformation, see table 1). Of course, more dedicated experiments of time-resolved material parameters (transient optical properties, temperatures and induced internal strains, etc.) and detailed calculations are desirable for improving knowledge and predictability of ablation outcomes in such ablation regime.

Finally, to highlight the interest of ultrashort pulse for ablation of metals, we define the energy specific ablation depth parameter: $\eta = d_{ablated}/E_{incident}$. The evolution of $\eta$ is plotted on Fig. 4 as a function of pulse duration and fluence and for all metals studied. For Cu and Ni, the highest energy specific ablation depth is obtained just above the threshold fluence ($F_{opt} \sim 2\text{-}3$ $F_{th}$). This is also measured for Al and W with a pulse duration of 15 fs. This result has already been shown for longer pulse durations. In [48] by considering the absorbed energy density evolving inside the sample along the z axis according to Beer-Lambert law, it is demonstrated that the highest energy specific ablation depth is achieved when $F/F_{th}=e$. The ablation threshold fluence being the same for durations ranging from 15 to 100 fs, the maximum energy specific ablation depth is achieved at the same fluence (as observed for most cases in figure 4). Our measurements extend this result to ultra-short pulses which confirms that there is no striking advantage in using pulses of extremely short duration (≤ 30 fs) for ablation of metal. Absence of significant differences in morphology of ablated craters (see Figure 6) at the two extreme pulse durations studied here (15 and 100 fs) supports this conclusion.

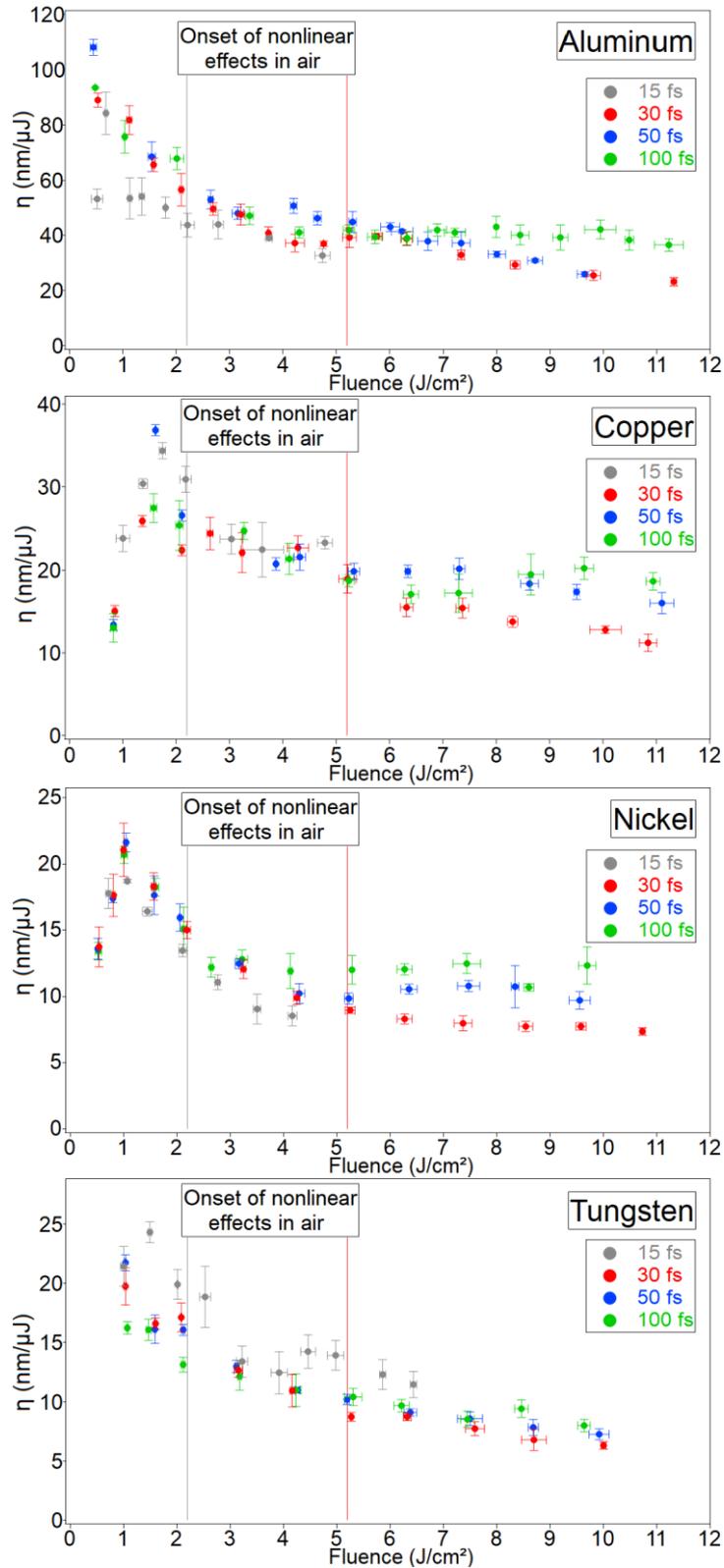

**Figure 4** Evolution of energy specific ablation depth parameter $\eta$ as a function of normalized fluence and for different pulse durations and for the four metals studied. Note that the difference of geometry of focusing does not allow comparing quantitatively the 15 fs case (Setup B) and the three other (30, 50 and 100 fs) cases (Setup A).

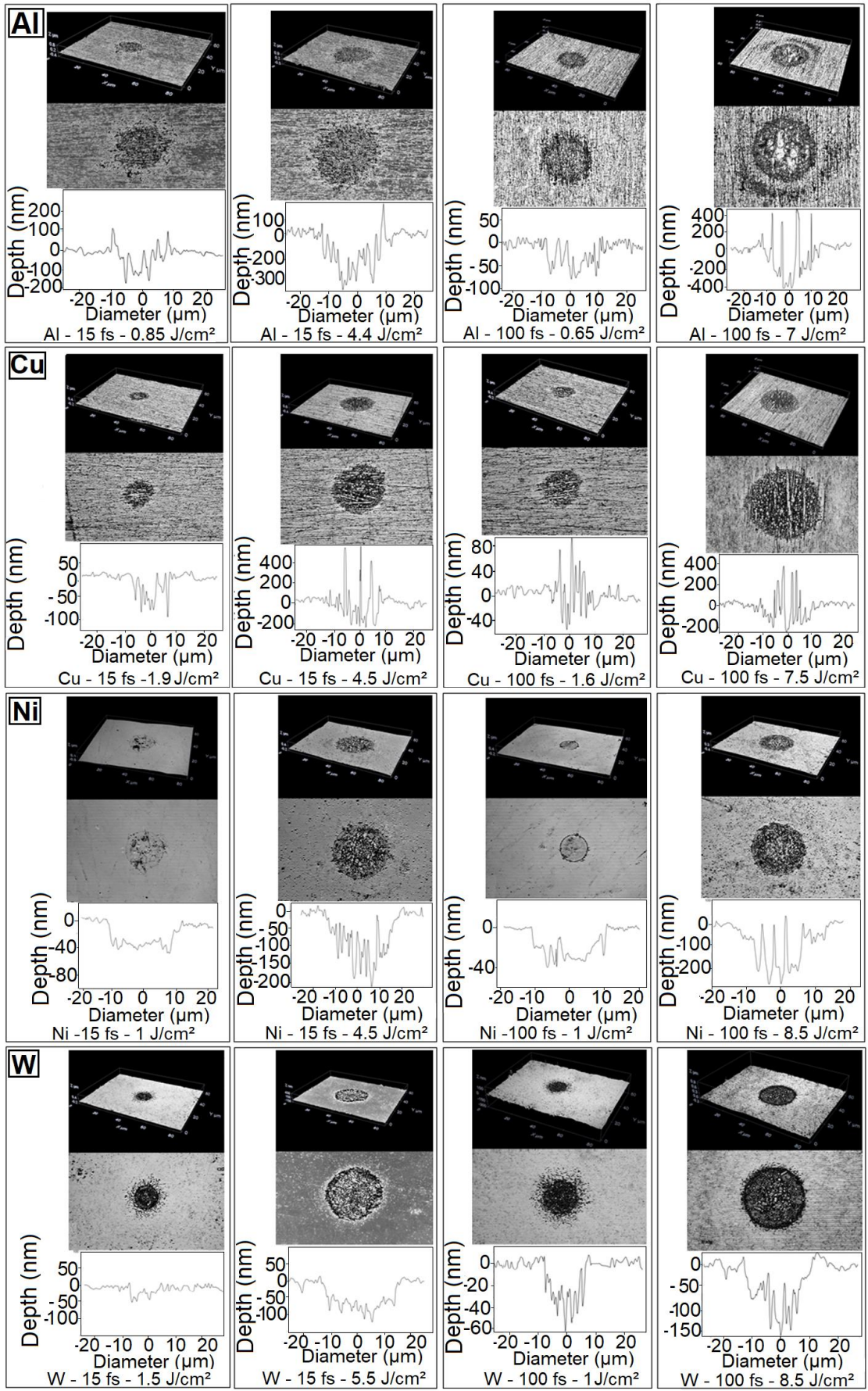

**Figure 5** Ablated crater morphology for the four metals at two pulse durations (15 and 100 fs) and for two fluences close and much superior to the ablation threshold $F_{th}$.

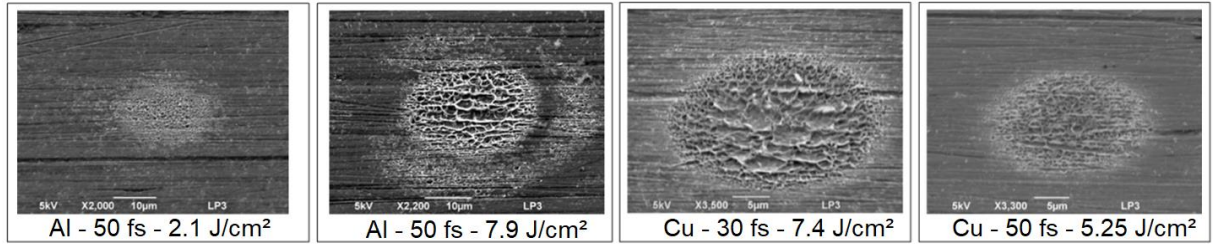

**Figure 6**: SEM images of Al and Cu to illustrate the surface and roughness of the resulting crater surface with evidence of occurrence of ablation thermal effects.

As a general comment, we observe a correlation between the initial surface finish of the samples (as represented by the *Ra* parameter, see table 1) and the resulting ablated surface morphology. Indeed, for metals having low *Ra* parameter (nickel and tungsten), the morphology of the ablation craters appears to be more regular than for the two other metals studied. Finally the resulting surface at the bottom of the crater, which is much superior to the initial roughness for all materials, shows evidence that strong thermal stress effects have developed following energy deposition (see for instance figure 6 for fine details for Al and Cu).

## 4 Conclusion

This paper brings an extended set of ablation data for four metals (Al, Cu, Ni and W) irradiated in air ambiance with single-shot femtosecond pulses of variable pulse duration (15 – 100 fs), so in a temporal range very poorly explored yet. We first provide the measurement of the laser-induced ablation threshold fluence for all metals showing constancy of this material characteristics over the pulsed duration range studied. Those data have been further confronted to ablation threshold scaling laws available from literature. Even if the non-dependent behavior of the ablation threshold is correctly predicted by the scaling laws (based on different electronic and thermal ablation scenarios), the quantitative agreement between prediction and measurement is fair for only three metals (Al, Ni and W). More detailed experimental and theoretical developments including detailed knowledge of the absorption coefficient and of the transient optical properties and further laser heating and relaxation in general are desirable for enhancing accuracy of threshold description and promote better predictability.

Moreover, we measured ablation characteristics (ablated depth and diameter) over an extended range of fluence (typically 1 – 10 J/cm²). Again, except at high fluences, the evolution of the ablation outcomes appears to be largely independent of the pulse duration suggesting the predominance of equilibrium processes (mainly thermal) in ablation scenario until extremely short pulse duration as it was often mentioned in literature for longer pulses [18,33,40]. Our results also highlight the interest to adjust the incident fluence just above the threshold (2-3 times) for optimal energy specific ablation depth. It is not surprising; indeed, when significantly increasing the incident fluence above the threshold, absorption tends to be higher [40,42,43] but the total heated volume is not dramatically increasing because of progressive reduced efficiency of energy transport at high electron excitation.

Finally, as another important outcome of this work, we show that there is no real interest in using pulses of few-optical-cycle pulse duration for ablation-based application processes. Indeed, ultrashort pulses do not provide any advantage in terms of energy specific ablation depth. Moreover, they are also difficult to handle experimentally because the propagation in air and optical devices can rapidly alter the spatio-temporal characteristics of the pulse. Correlatively, the ablation threshold fluence being constant until pulse duration of few-optical-cycle, the absence of any peculiar vulnerability of metal-based optical components (like aluminum mirrors) to femtosecond laser exposition is also a valuable outcome of our research for femtosecond laser technology and for the involved scientific community.

**Funding**

Financial support of the ASUR platform was provided by the European Community, Ministry of Research and High Education, Region Provence-Alpes-Côte d'Azur, Department of Bouches-du-Rhône, City of Marseille, CNRS, and Aix-Marseille University.

**Acknowledgments**

Thibault Genieys acknowledges the support of DGA – Direction Générale de l'Armement (Ministry of Defense) and Aix-Marseille University for his Ph'D grant.

**References**

1. K. M. T. Ahmmed, C. Grambow and A.-M. Kietzig, Micromachines 5, 1219 (2014).

2. J. Cheng, C.-S. Liu, S. Shang, D. Liu, W. Perrie, G. Dearden, K. Watkins, Optics and Laser Technology 46, 88 (2013).

3. M. Lenzner, J. Krüger, S. Sartania, Z. Cheng, Ch. Spielmann, G. Mourou, W. Kautek, F. Krausz, Phys. Rev. Lett. 80 (18), 4076 (1998).

4. C.S.R. Nathala, A. Ajami, W. Husinsky, B. Farooq, S.I. Kudryashov, A. Daskalova, I. Bliznakova, A. Assion, Appl. Phys. A 122, 107 (2016).

5. M.E. Shaheen, J.E. Gagnon, J.B. Fryer, Laser Phys. 24, 106102 (2014).

6. P.T. Mannion, J. Magee, E. Coyne, G.M. O'Connor, T.J. Glynn, Appl. Surf. Sci. 233, 275 (2004).

7. B. Chimier, O. Utéza, N. Sanner, M. Sentis, T. Itina, P. Lassonde, F. Légaré, F. Vidal and J. C. Kieffer, Phys. Rev. B 84, 094104 (2011).

8. O. Utéza, N. Sanner, B. Chimier, A. Brocas, N. Varkentina, M. Sentis, P. Lassonde, F. Légaré, J.C. Kieffer, Appl. Phys. A 105, 131 (2011).

9. J. Bonse, S. Baudach, J. Krüger, W. Kautek, M. Lenzner, Appl. Phys. A 74, 19 (2002).

10. P.P. Pronko, S.K. Dutta, D. Du, R.K. Singh, J. Appl. Phys. 78 (10), 6233 (1995).

11. B.C. Stuart, M.D. Feit, S. Herman, A.M. Rubenchik, B.W. Shore, M.D. Perry, J. Opt. Soc. Am. B 13 (2), 459 (1996).

12. S.E. Kirkwood, A.C. Van Popta, Y.Y. Tsui, R. Fedosejevs, Appl. Phys. A 81, 729 (2005).

13. S. Martin, J. Krüger, A. Hertwig, A. Fiedler, W. Kautek, Femtosecond laser interaction with protection materials, Appl. Surf. Sci. 208/209, 333 (2003).

14. J. Byskov-Nielsen, J.-M. Savolainen, M. Snogdahl Christensen, P. Balling, Appl. Phys. A 101, 97 (2010).

15. G. Mourou, T. Tajima, Science 331, 41 (2011).

16. S. Mishra, V. Yadava, Optics and Lasers in Engineering 73, 89 (2015).

17. S.Y. Wang, Y. Ren, C.W. Cheng, J.K. Chen, D.Y. Tzou, Appl. Surf. Sci. 265, 302 (2013).

18. C.W. Cheng, S.Y. Wang, K.P. Chang, J.K. Chen, Appl. Surf. Sci. 361, 41 (2016).

19. C. Pasquier, P. Blandin, R. Clady, N. Sanner, M. Sentis, O. Utéza, Yu Li, Shen Yan long, Opt. Comm. 355, 230 (2015).

20. "Laser heating of metals", Ed. by A.M. Prokhorov, V.I. Konov, I. Ursu, I.N. Mihailescu (Adam-Hilger, Bristol, England,1990).

21. A. D. Rakić, A. B. Djurišic, J. M. Elazar, and M. L. Majewski, Appl. Opt. 37, 5271 (1998).

22. "Handbook of Optical Materials", Ed. by M.J. Weber (CRC Press, LLC, NYC, USA,2003).

23. "CRC Handbook of Chemistry and Physics, 84$^{th}$ edn." Ed. by D.R. Lide (CRC Press, London, 2003–2004).

24. N.W. Ashcroft, N.D. Mermin, "Physique des Solides", (EDP Sciences, Les Ulis, 2002).

25. P.B. Johnson and R.W. Christy, Phys. Rev. B 6 (12), 4370 (1972).

26. M.A. Ordal, R.J. Bell, R.W. Alexander, L.L. Long, and M.R. Querry, Appl. Opt. 24 (24), 4493 (1985).

27. J.M. Liu, Opt. Lett. **7**, 196 (1982).

28. N. Sanner, O. Utéza, B. Bussiere, G. Coustillier, A. Leray, T. Itina, M. Sentis, Appl. Phys. A 94, 889 (2009).

29. P.B. Corkum, F. Brundel, N.K. Sherman, T. Srinivasanrao, Phys. Rev. Lett. 61 (25), 2886 (1988).

30. E.G Gamaly, A.V. Rode, B. Luther-Davies, V.T. Tikhonchuk, Phys. Plasmas 9, 949 (2002).

31. C. Momma, S. Nolte, B. N. Chichkov, F. v. Alvensleben, A. Tünnermann, Appl. Surf. Sci. 109/110, 15 (1997).

32. A. Miotello, R. Kelly, Appl. Phys. A 69, S67-S73 (1999).

33. L.V. Zhigilei, Z. Lin and D.S. Ivanov, J. Phys. Chem. C 113, 11892 (2009).

34. C. Pasquier, M. Sentis, O. Utéza, N. Sanner, Appl. Phys. Lett. 109 (5), 051102 (2016).

35. S. Akturk, C. D'Amico, M. Franco, A. Couairon and A. Mysyrowicz, Opt. Express 15 (23), 15260 (2007).

36. S. Nolte, C. Momma, H. Jacobs, A. Tunnermann, B. N. Chichkov, B. Wellegehausen, H. Welling, J. Opt. Soc. Am. B 14 (10), 2716 (1997).

37. R. Le Harzic, D. Breitling, M. Weikert, S. Sommer, C. Föhl, F. Dausinger, S. Valette, C. Donnet, E. Audouard, Appl. Phys. A 80, 1589 (2005).

38. Z. Lin, L.V. Zhigilei, V. Celli, Phys. Rev. B. 77, 075133 (2008).

39. A. Suslova, A. Hassanein, Laser and Particle Beams, 35 (3), 415 (2017).

40. S.E. Kirkwood, Y.Y. Tsui, R. Fedosejevs, A.V. Brantov, V.Yu. Bychenkov, Phys. Rev. B 79, 144120 (2009).

41. B.H. Christensen, K. Vestentoft, P. Balling, Appl. Surf. Sci. 253, 6347 (2007).

42. T. Genieys, M. Sentis, O. Utéza, Appl. Phys. A 126, 263 (2020).

43. T. Genieys, "Ablation laser en régime ultracourt de cibles diélectriques et métalliques », Ph.D. thesis (Aix-Marseille University, 2019), http://www.theses.fr

44. J. Hohlfeld, S.-S. Wellershoff, J. Güdde, U. Conrad, V. Jähnke, E. Matthias, Chem. Phys. 251, 237 (2000).

45. E. Knoesel, A. Hotzel, and M. Wolf, Phys. Rev. B 57 (20), 12812 (1998).

46. X.Y. Wang, D.M. Riffe, Y.-S Lee, and M.C. Downer, Phys. Rev. B 50 (11), 8016 (1994).

47. N.M. Bulgakova, A.V. Bulgakov, Appl. Phys. A 73, 199 (2001).

48. B. Neuenschwander, B. Jaeggi, M. Schmid, V. Rouffiange, P.-E. Martin, Proc. of SPIE Vol. 8243 824307-1 (2012), https://doi.org/10.1117/12.908583